# Graphene Oxide vs. Reduced Graphene Oxide as saturable absorbers for Er-doped passively mode-locked fiber laser

Grzegorz Sobon,[1,*] Jaroslaw Sotor,[1] Joanna Jagiello,[2] Rafal Kozinski,[2] Mariusz Zdrojek,[3] Marcin Holdynski,[4] Piotr Paletko,[5] Jakub Boguslawski,[1] Ludwika Lipinska,[2] and Krzysztof M. Abramski[1]

[1]*Laser & Fiber Electronics Group, Wroclaw University of Technology,*
*Wybrzeze Wyspianskiego 27,50-370 Wroclaw, Poland*

[2]*Institute of Electronic Materials Technology, Wolczynska 133, 01-919 Warsaw, Poland*

[3]*Faculty of Physics, Warsaw University of Technology, Koszykowa 75, 00-662 Warsaw, Poland*

[4]*Institute of Physical Chemistry, Polish Academy of Sciences, Kasprzaka 44/52, 01-224 Warsaw, Poland*

[5]*Faculty of Microsystem Electronics and Photonics, Wroclaw University of Technology, Janiszewskiego 11/17, Wroclaw 50-372, Poland*

[*]*grzegorz.sobon@pwr.wroc.pl*

**Abstract:** In this work we demonstrate comprehensive studies on graphene oxide (GO) and reduced graphene oxide (rGO) based saturable absorbers (SA) for mode-locking of Er-doped fiber lasers. The paper describes the fabrication process of both saturable absorbers and detailed comparison of their parameters. Our results show, that there is no significant difference in the laser performance between the investigated SA. Both provided stable, mode-locked operation with sub-400 fs soliton pulses and more than 9 nm optical bandwidth at 1560 nm center wavelength. It has been shown that GO might be successfully used as an efficient SA without the need of its reduction to rGO. Taking into account simpler manufacturing technology and the possibility of mass production, GO seems to be a good candidate as a cost-effective material for saturable absorbers for Er-doped fiber lasers.

**1. Introduction**

Ultra-short pulse fiber lasers operating in the near infrared region have found many applications in industry and basic science such as laser micromachining [1], micro-surgery [2], optical metrology [3], optical imaging [4] and THz generation [5]. In 2009, Bao et al. and Hasan et al. have demonstrated a new technique of generating ultra-short pulses from passively mode-locked (PML) fiber lasers by incorporating graphene-based saturable absorbers [6,7]. Since then, various setups have been proposed, utilizing mainly erbium-doped fibers [8-22]. So far, many different techniques of obtaining graphene for PML have been used. The most common are: chemical vapor deposition (CVD), chemical functionalization and mechanical exfoliation. The CVD-graphene may be grown on various metallic substrates (e.g. Ni, Cu) and afterwards easily transferred onto fiber connectors forming a saturable absorber [6,8-11]. CVD technique allows to precisely control the number of grown layers and, in consequence, tune the optical parameters of the SA [6,11]. Graphene flakes might also be obtained by direct exfoliation of natural graphite by ultrasonification it in various organic solvents e.g. dimethyloformamide (DMF) [12]. Obtained graphene suspension is cast onto flat fiber connector by optical deposition. Graphite can be also exfoliated in aqueous solutions with surfactant e.g. sodium deoxycholate. Then the water suspension containing graphene flakes is mixed with polyvinyl alcohol (PVA) [13-15]. Such homogenous polymer solutions may also be applied to the optical substrates or fiber connectors and slowly dried to obtain an uniform composite layer. Mechanical exfoliation (or so called "scotch tape method") is obviously the easiest method of obtaining graphene flakes useful for mode-locking of fiber lasers [16-18]. Although, mechanical exfoliation does not allow to control the layer thickness and the repeatability of the process is very poor.

Very often graphene flakes (rGO) for saturable absorbers are obtained by chemical reduction of so called graphene oxide (GO) [19-21]. Also, for this purpose polymer composite e. g. polyvinylidene fluoride (PVDF)/rGO can be used [22]. GO is an atomically thin sheet of carbon covalently bonded with functional groups containing oxygen. Therefore, it contains $sp^2$ and $sp^3$ hybridized carbon atoms. Usually it is used as a precursor for rGO, but recently it has attracted much attention due to its optical properties [23]. It has been shown, that GO, similar to rGO, exhibits saturable absorption [24] which makes it suitable for passive mode-locking of lasers. Femtosecond pulse generation with the usage of GO saturable absorbers was recently demonstrated by Bonaccorso et al. [25], Liu et al. [26] and Xu et al. [27]. The pulse duration obtained by Xu et al. (200 fs) is comparable to that obtained with flake-graphene solution [12]. Promising results show, that GO can successfully compete with graphene as saturable absorber.

Nevertheless, it is currently impossible to objectively compare the performance of fiber lasers mode-locked with GO and graphene (e.g. rGO), since there are no direct comparisons of both saturable absorbers. In our work we perform a detailed comparison between two types of SA: based on graphene oxide and reduced graphene oxide. In order to provide a reliable comparison, both saturable absorbers were tested in the same laboratory conditions (the same laser resonator). The GO and rGO layers were deposited on fused silica windows and placed inside the cavity, forming a free-space coupled, transmission saturable absorber. The results show, that there is almost no difference between the radiation parameters achieved with GO and rGO. The optical bandwidth and pulse duration remain unchanged. It may suggest that the presence of $sp^3$ hybridized carbons in the lattice does not completely destroy the nonlinear optical properties of graphene. In consequence it may not be necessary to perform a complicated GO reduction process in order to fabricate an efficient saturable absorber.

## 2. GO and rGO preparation and characterization

*2.1 Sample preparation*

Graphene oxide (GO) was prepared through a modified Hummers method from expanded acid washed graphite flakes [28]. The synthesis involves the following steps. First, 5 g of graphite was added into 125 ml of $H_2SO_4$. Next, 2.75 g of $NaNO_3$ was added before start of the reaction. Subsequently, the beaker with reagents was put into water/ice bath in order to keep it below 5°C. 15 g of $KMnO_4$ was added in portions into the mixture, which was vigorously stirred. After addition of the oxidant, the beaker was heated and kept at 30-35°C with continuous stirring. Afterward it was left at room temperature for overnight. In the next step, the deionized water was added so that the temperature did not exceed 35°C. The beaker was put into a water bath at a temperature of 35°C and stirred. The mixture was then heated to 95°C and kept under these conditions for 15 min. To stop the reaction 280 ml of deionized water and 5 ml of $H_2O_2$ were added. The mixture was then rinsed with HCl solution to remove the sulfate ions and with deionized water in order to remove chloride ions.

Reduced graphene oxide (rGO) was obtained with the use of benzylamine as reducing and functionalizing agent [29] and sodium borohydride as reducing agent [30,31]. Such combination of reductors was used for the first time, to the best of our knowledge. 25 mg of GO were dispersed in 150 ml of deionized water, followed by addition of 3 ml of benzylamine. The mixture was heated to 90°C and maintained at this temperature over 2 hours with simultaneous stirring on a magnetic stirrer. Such obtained black dispersion was rinsed three times with deionized water by centrifugation. The precipitation was dispersed in deionized water by ultrasonicator. After this procedure 1.5 g of $NaBH_4$ and 1 g of KOH were added into the suspension and it was heated to 70°C with simultaneous stirring over a 2-hours period. The obtained rGO was rinsed as previously and dispersed in 100 ml of N-methylpyrrolidone by ultrasonication to prevent rGO flakes aggregation [32].

In order to characterize the chemical parameters (by SEM, XPS and Raman spectroscopy) of GO and rGO it was deposited from suspension on $Si/SiO_2$ substrates by immersion and dried at 80°C. To characterize its optical parameters and investigate the morphology by atomic force microscopy (AFM), GO and rGO were deposited on the fused silica windows using self-made applicator and dried slowly at 40°C. Photographs of the ½-inch plates are shown in Fig. 1.

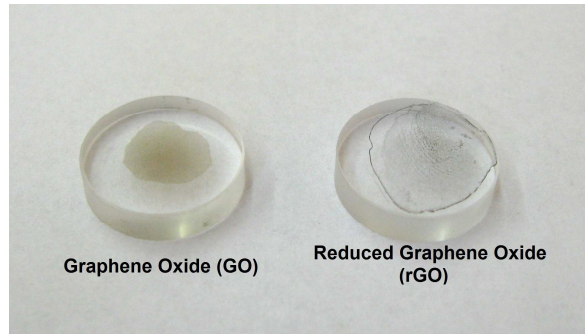

Fig. 1. Photograph of the ½-inch fused silica windows with GO and rGO layers.

*2.2 Sample characterization*

Morphological properties were investigated with scanning electron microscopy (SEM), using Auriga CrossBeam Workstation (Carl Zeiss). AFM was acquired in tapping mode using Veeco Nanoman V microscope with Bruker MPP-11100-10 silica probe. Elemental composition analysis was carried out by X-ray photoelectron spectroscopy (Microlab 350 spectrometer) using $Mg_{Ka}$ non-monochromated radiation (1253.6 eV, 200 W) as the excitation source. The XPS spectra were fitted using the Advantage (Thermo Scientific) software. The microstructure was characterized by Raman spectroscopy (Dilor XY-800 spectrometer), using 514 nm wavelength of an argon-ion laser.

Fig. 2 shows SEM images of GO and chemically reduced GO. Single flakes of GO may be observed. Graphene oxide flakes have relatively large surface (with the edge of sheets about the size of micrometers) and its morphology resembles thin curtain. These parameters indicate very good exfoliation of graphite during oxidation process. The rGO solution had a higher concentration than that of GO and therefore the sheets applied on a $Si/SiO_2$ substrate overlap more and seem to form a compact structure. The surface morphology resembles strongly folded curtain, what indicates that rGO flakes are overlapped rather than aggregated.

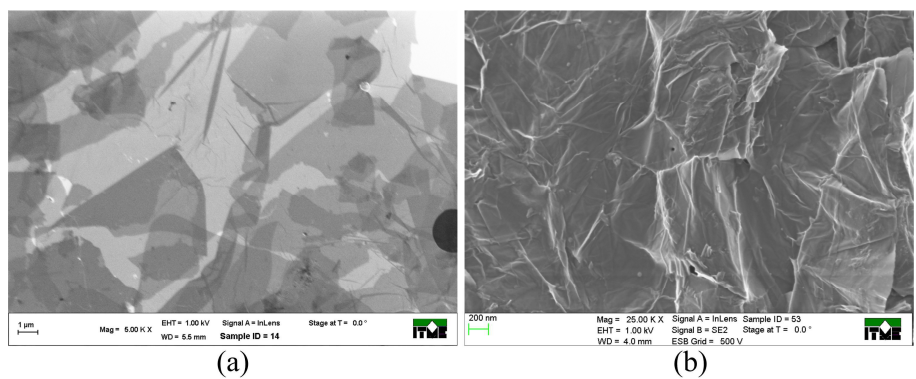

Fig. 2. SEM images of GO (a) and reduced GO (b).

Raman spectroscopy is widely used to characterize crystal structure, disorder and defects in graphene-based materials [33]. For example, the reduction process of GO can manifest itself in Raman spectra by the changes in relative intensity of two main peaks: D and G [34]. We use this information to verify the reduction process. Fig. 2 shows the Raman spectra of GO and reduced GO. The D peak of GO located at 1352 cm$^{-1}$ and at 1350 cm$^{-1}$ for rGO streams from a defect-induced breathing mode of sp$^2$ rings [29]. It is common to all sp$^2$ carbon lattice and arises from the stretching of C-C bond. The G peak at around 1600 cm$^{-1}$ for GO and at 1599 cm$^{-1}$ for rGO is due to the first order scattering of the E2g phonon of sp$^2$ C atoms [29]. The intensity of the D band is related to the size of the in-plane sp$^2$ domains [33]. The increase of the D peak intensity indicates forming more sp$^2$ domains. The relative intensity ratio of both peaks ($I_D/I_G$) is a measure of disorder degree and is inversely proportional to the average size of the sp$^2$ clusters [33,35]. As it is seen in Fig. 3., the D/G intensity ratio for rGO is larger than that for GO (1.70 for rGO and 1.21 for GO). This suggest that new (or more) graphitic domains are formed and the sp$^2$ cluster number is increased [29] after process described above, showing good reduction efficiency of benzylamine and sodium borohydride.

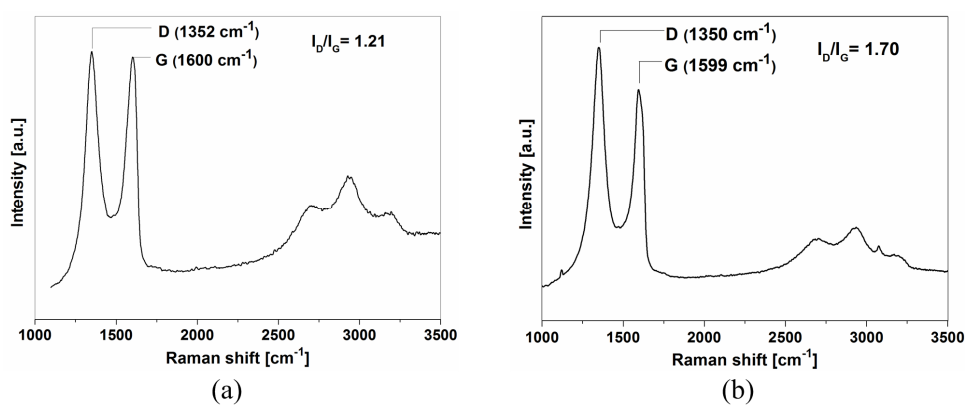

Fig. 3. Raman spectra of GO (a) and reduced GO (b). Both Raman spectra were recorded with 514 nm laser line and with the low laser power (1 mW)

X-ray photoelectron spectroscopy (XPS) was employed to analyze the structure of graphene oxide and reduced graphene oxide and also to confirm the results obtained by the Raman spectroscopy. It is a surface-sensitive analytical technique that is useful to determine the chemical environment of atoms, in this case of carbon atoms in rGO structure. Fig. 4 shows C1s XPS spectra of GO and Fig. 5 presents the curve-fitted C1s of GO reduced with the use of benzylamine and sodium borohydride (respectively). Table 1 provides an analysis of the spectrum peaks: the most probable origin of the peaks with their binding energies and atomic percentage of each group.

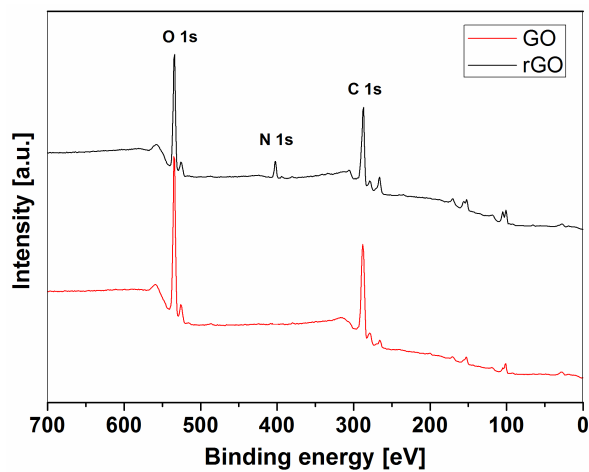

Fig. 4. XPS spectra of GO and reduced GO

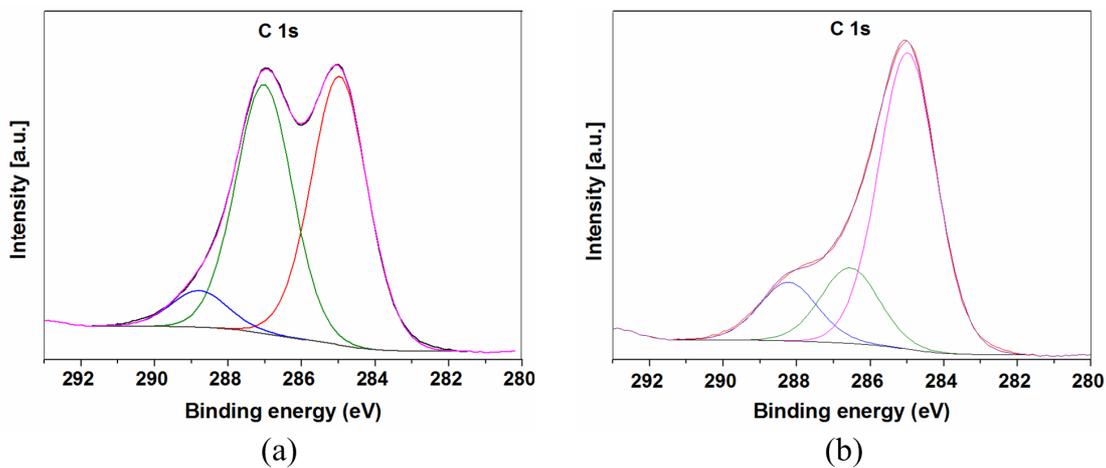

Fig. 5. C 1s XPS spectra of (a) GO and (b) reduced GO

Table 1. XPS data of GO

| GO | Peak BE (eV) | At. % | Bond |
|---|---|---|---|
| C1s | 285.0 | 48.6 | C – C and C=C |
| C1s | 287.0 | 44.7 | C – O (epoxy, hydroxyl groups) |
| C1s | 288.2 | 6.7 | C = O (carbonyl groups) |

Table 2. XPS data of rGO

| rGO | Peak BE (eV) | At. % | Bond |
|---|---|---|---|
| C1s | 285.0 | 69.0 | C – C and C=C |
| C1s | 286.6 | 17.5 | C-O (hydroxyl, epoxy groups), C-N |
| C1s | 288.2 | 13.5 | C = O (carbony groups) |

All spectra were calibrated to the position of the C-C peak of $285.0 \pm 0.2$ eV. The resolution of XPS spectrometer does not allow to analyze peaks of C-C and C=C separately, therefore it will be treated as a single signal and compared with peaks corresponding to the carbon atoms bounded with other groups. C1s XPS spectra of GO indicates considerably degree of oxidation that means the presence of different oxygen functional groups in GO structure (e.g. carbonyl, epoxy, hydroxyl groups) [31,36]. Peaks corresponding to the covalent bonds of carbon and oxygen atoms are more intense for GO than for rGO. The CC/CO intensity ratio of GO is much lower (0.95) than CC/CO ratio of reduced GO (2.23), where "CC" refers to the sum of C-C and C=C bonds and "CO" applies to all combinations of carbon and oxygen atoms bonds. Moreover, "CC" intensity of rGO was calculated using the sum of C-N and C-O bonds, because it was not possible to separate C-O and C-N peaks by our spectrometer. After the separation of the peaks the ratio might be higher. These results suggest significant removal of oxygen functional groups. After reduction process the new peak (400 eV) associated with N 1s (Fig. 4) and the C-N band (Fig. 5b) was created. This reflects the simultaneous functionalization of GO with benzylamine [29].

The AFM scan images of GO and rGO are presented in Fig. 6a and Fig 6b, respectively. In the case of GO, the surface is relatively rough, since the flakes are forming irregular stacks. Mono- and bi-layer flakes may be seen on the rGO sample with few µm length and 1-2 nm height. Due to local contaminations on the sample, the investigated area contained some peaks with height at the level of 150 nm. In order to improve the readability of the rGO picture, those peaks were scaled down to 10 nm, causing white spots on the image.

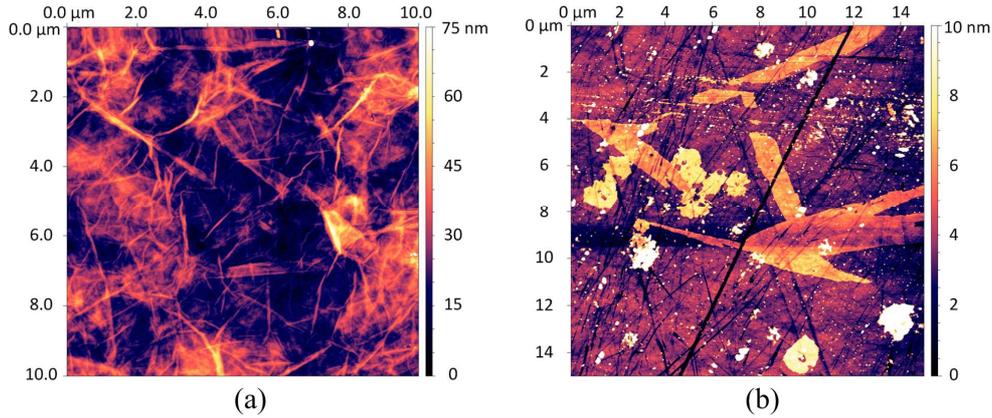

Fig. 6. AFM scan image of the GO (a) and rGO (b) surface.

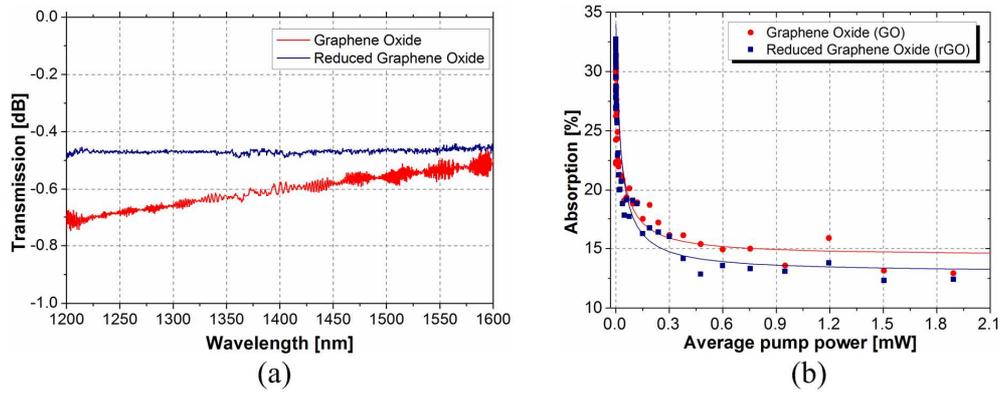

Fig. 7. Measured linear transmission spectrum (a) and power-dependent transmission (b) of the GO and rGO

Fig. 7a shows the measured transmission of GO and rGO in the spectral range from 1200 to 1600 nm, performed using a broadband white light source and an optical spectrum analyzer. It can be seen, that the rGO sample absorbs around 12% of incident light (average loss at the level of 0.48 dB). The transmission of GO is not flat in the measured range and the losses vary from 0.7 to 0.55 dB. The wavelength-dependent transmission is a typical feature of GO and was already observed [37,38]. It might be caused by the presence of functional groups containing oxygen, which may absorb shorter wavelengths more likely. In order to determine the modulation depth of the saturable absorbers, we have performed a power-dependent transmission measurement in a setup similar to that in Ref. [2], using a 169 MHz mode-locked laser with 150 fs pulse duration (1560 nm) and 2.1 mW average output power as a pump source. The signal was directed to the SA through a variable optical attenuator (VOA). The results of the power-dependent transmission are plotted in Fig. 7b. The measurement results may be fitted with a two-level saturable absorber model curve, given by:

$$\alpha(I) = \frac{\alpha_0}{1+\dfrac{I}{I_{sat}}} + \alpha_{ns}, \qquad (1)$$

where $\alpha(I)$ is the absorption coefficient, $I$ is the light intensity, $I_{sat}$ is the saturation intensity, $\alpha_0$ is the modulation depth and $\alpha_{ns}$ denotes the non-saturable losses. Based on the results, the modulation depths of rGO and GO are 21% and 18%, respectively.

## 3. Laser setup and results

The experimental setup of the mode-locked laser is presented in Fig. 8. The resonator consists of a 30 cm long highly-erbium doped fiber (nLight Liekki Er110), a fiber isolator, 980/1550 single-mode WDM coupler, in-line fiber polarization controller and a 10% output coupler. The saturable absorber (GO or rGO) on fused silica substrate is inserted between two GRIN-lens collimators (200 mm working distance). It was placed on a three-axis positioning stage in order to adjust the position of the sample.

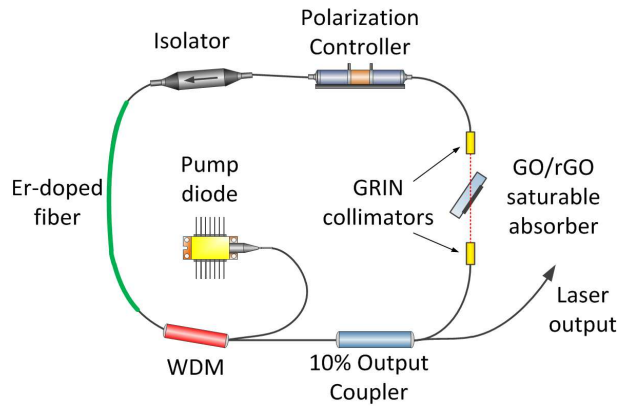

Fig. 8. Experimental setup of the mode-locked laser

The laser is counter-directionally pumped by a 980 nm laser diode (Oclaro LC96U). The total length of the fibers in the resonator is approx. 3.1 m. There are only two types of fibers used in the cavity: erbium-doped fiber (EDF) with positive group velocity dispersion (GVD): 0.012 ps$^2$/m and single mode fiber (SMF) with -0.022 ps$^2$/m GVD, thus, the total net dispersion is anomalous (-0.06 ps$^2$). The laser performance was observed using optical spectrum analyzer (Yokogawa AQ6370B), 350 MHz digital oscilloscope (Hameg HMO3524), 7 GHz RF spectrum analyzer (Agilent EXA N9010A) coupled with a 30 GHz photodetector (OptiLab PD-30), and an interferometric autocorrelator.

After launching the pump diode at low power (below 30 mW), the laser starts to operate in the CW regime. The threshold pump power for mode-locking was 37 mW and 33 mW for the GO and rGO, respectively. In order to provide best stability and performance, all measurements of the GO laser were done with 95 mW (GO) and 82 mW (rGO) pumping power. The fused silica plates used as substrates for both SA were uncoated. Hence, when the SA was inserted inside the laser resonator perpendicular to the optical axis, parasitic etalon effect was observed and mode-locking could not start due to strong spectral filtering. In order to eliminate the back-reflections and enable mode-locking, the SA was placed near the Brewster angle. As a result, the etalon effect was eliminated and pulsed operation was achieved after a slight alignment of polarization controller. Since the Brewster-angled fused silica plate acts as a weak polarizer, the nonlinear polarization rotation mechanism might be supported in such configuration. In order to confirm, that the mode-locking is an effect of the saturable absorption in GO/rGO, not nonlinear polarization rotation, we have firstly inserted a clean fused silica window (without graphene). No signs of mode-locking were observed at all possible positions of the polarization controller. During operation the SA was moved in all three X-Y-Z axes without any mode-locking perturbations. The pulse operation was lost when the optical beam was positioned outside the GO and rGO layers (on the clean area of the plates).

The recorded optical spectra of the laser are depicted in Fig. 9. Both have comparable, soliton-like shapes with visible Kelly's sidebands and full width at half maximum (FWHM) bandwidth at the level of 9.3 nm (GO) and 9.2 nm (rGO). The GO spectrum is slightly blue-shifted due to higher insertion losses of the saturable absorber. Both do not have any signs of parasitic CW lasing.

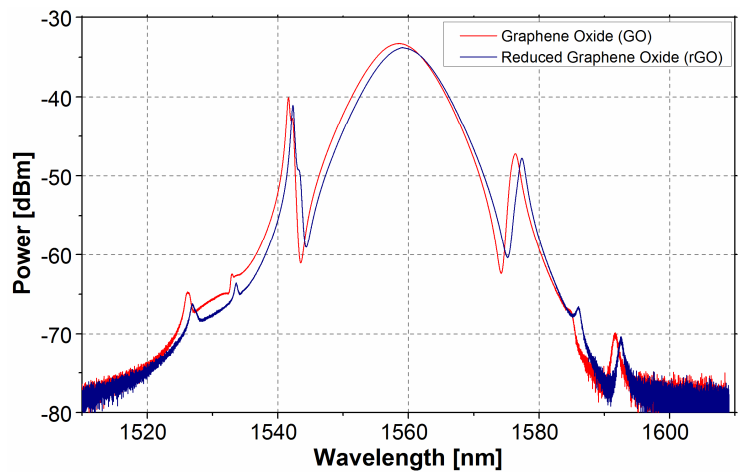

Fig. 9. Comparison of the optical spectra with GO (red line) and RGO (blue line)

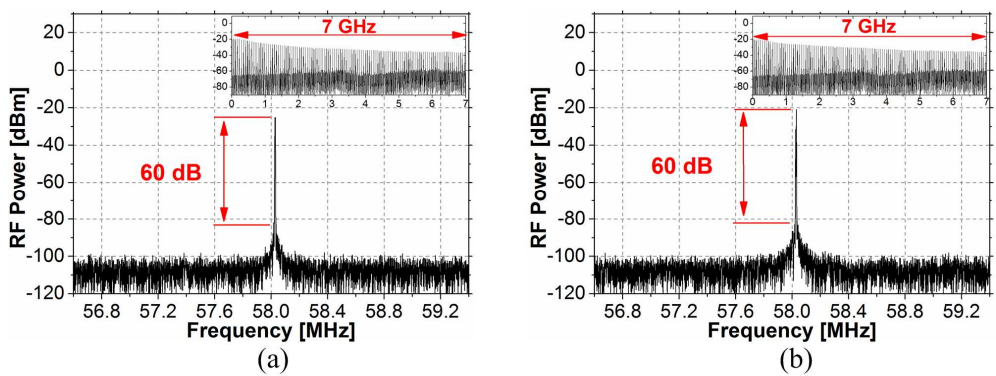

Fig. 10. RF spectra of the laser with GO (a) and rGO (b) recorded with 3 MHz span and 620 Hz RBW.

Fig. 10 illustrates the RF spectrum of the laser measured with 620 Hz resolution bandwidth (RBW) and 3 MHz span. The signal to noise ratio (SNR) is at the level of 60 dB in both cases and the differences between two samples are imperceptible. The measured repetition frequency of the resonator is 58 MHz. The broad spectrum of harmonics (with 7 GHz span) generated by the laser is presented inset the graphs in Fig. 10.

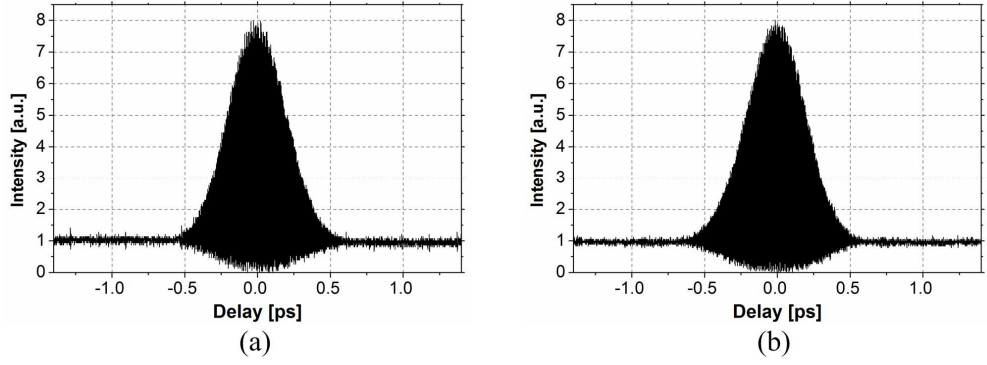

Fig. 11. Autocorrelation traces of the 390 fs pulses obtained with GO (a) and rGO (b)

The autocorrelation traces of the output pulses are depicted in Fig. 11. In both cases the pulse width is the same and is equal to 390 fs (assuming a sech$^2$ pulse shape, typical for soliton lasers). No signs of pulse pair generation or pulse breaking were observed. The calculated Time-Bandwidth Products (TBP) are 0.448 and 0.442 for GO and rGO, respectively.

Assuming the fiber nonlinear coefficient $\gamma=3$ W$^{-1}$km$^{-1}$ and the average laser resonator dispersion $\beta_2=-20$ps$^2$/km, we may determine the soliton order $N$ using the formula given by [39]:

$$N = \sqrt{\frac{\gamma \cdot P \cdot \tau^2}{|\beta_2|}}, \qquad (2)$$

where $P$ denotes the pulse peak power (see Table 3) and $\tau$ is the FWHM pulse duration divided by 1.762. The $N$ value were at the level of 0.79 and 0.75 for laser with GO and rGO, respectively. Both of them meet the stability criteria 1.5>N>0.5 for fundamental solitons [39]. The comparison between all achieved laser parameters are summarized in Table 3.

Table 3. Summary of the laser parameters with GO and rGO

| Parameter | Value | |
|---|---|---|
| | Graphene Oxide (GO) | Reduced Graphene Oxide (rGO) |
| FWHM bandwidth | 9.3 nm | 9.2 nm |
| Pulse duration | 390 fs | 390 fs |
| TBP | 0.448 | 0.442 |
| Pulse Energy | 33.7 pJ | 29.8 pJ |
| Peak Power | 86.4 W | 76.4 W |
| Soliton Order N | 0.79 | 0.75 |
| RF SNR | 60 dB | 60 dB |
| Pump power | 92 mW | 82 mW |
| Output power | 1.96 mW | 1.68 mW |
| Center wavelength | 1558 nm | 1559 nm |

The performance of the mode-locking with both investigated saturable absorbers is comparable to previous reports on soliton fiber lasers, even with those on mono- or bi-layer graphene grown by CVD [1,6,8]. The shortest pulse reported so far from an all-anomalous dispersion cavity is 415 fs [8], achieved with mono-layer graphene SA with over 60% modulation depth. We believe, that in our setup the pulse duration might be reduced by managing the total dispersion (achieving a near-zero dispersion cavity).

**4. Summary and conclusions**

Summarizing, we have demonstrated comparative experiments with Er-doped fiber laser mode-locked by two saturable absorbers: based on graphene oxide and reduced graphene oxide. Prepared absorbers were fully characterized using: Raman spectroscopy, XPS, AFM and absorption measurements. Obtained results confirm that used oxidation and reduction method provide saturable absorbers with relatively high modulation depth (21% and 18% for rGO and GO, respectively) and low non-saturable losses (around 15% for both SA). Both absorbers supported mode-locked operation in presented laser configuration. Lasers generated soliton pulses with 390 fs duration with over 9 nm of FWHM bandwidth. There were no significant differences between laser parameters obtained in both investigated set-ups. Described technology of oxidation of graphite to graphene oxide is a reproducible method, that allows to obtain material with well-defined structure. GO is a semi-product in the synthesis of rGO and therefore GO is easier and faster to obtain than graphene. GO forms stable aqueous dispersion and does not require the use of any organic compounds- it can be directly applied on the fused silica windows. Reduced graphene oxide flakes (as well as exfoliated graphene) because of its hydrophobic properties, aggregates in water and must be transferred to the proper, usually toxic organic solution. Beyond toxicity, the price of such prepared rGO increases significantly. Comparing the scale to the cost of production and including comparable effect of GO and rGO in the use as saturable absorber for fiber laser, GO seems to be the most promising material.

**Acknowledgements**

Work presented in this paper was supported by the National Science Centre (NCN, Poland) under the project "Saturable absorption in atomic-layer graphene for ultrashort pulse generation in fiber lasers" (decision no. DEC-2011/03/B/ST7/00208). We acknowledge the support of Zygmunt Luczynski and Zdzislaw Jankiewicz (Institute of Electronic Materials Technology, Warsaw), who inspired the groups from Warsaw and Wroclaw to collaborate and Pawel Kaczmarek (Laser & Fiber Electronics Group, Wroclaw) for helpful discussions. The AFM measurements were done at the Faculty of Microsystem Electronics and Photonics by the courtesy of Teodor Gotszalk.